# Doping Influence on $Sm_{1-x}Th_xOFeAs$ Superconducting Properties: Observation of Intrinsic Multiple Andreev Reflections Effect and Superconducting Parameters Determination

T. E. Kuzmicheva[a,b], S. A. Kuzmichev[b], N. D. Zhigadlo[c]

[a] *Lebedev Physical Institute RAS, 119991 Moscow, Russia*
*e-mail: kute@sci.lebedev.ru*
[b] *Lomonosov Moscow State University, Faculty of Physics, 119991 Moscow, Russia*
[c] *Laboratory for Solid State Physics, ETH Zurich, CH-8093 Zurich, Switzerland*

We studied SNS- and S-N-S-N-…-S contacts (where S – superconductor, N – normal metal) formed by "break-junction" technique in polycrystalline $Sm_{1-x}Th_xOFeAs$ superconductor samples with critical temperatures $T_C = (34 - 45)$ K. In such contacts (intrinsic) multiple Andreev reflections effects were observed. Using spectroscopies based on these effects, we detected two independent bulk order parameters and determined their magnitudes. Theoretical analysis of the large and the small gap temperature dependences revealed superconducting properties of $Sm_{1-x}Th_xOFeAs$ to be driven by intraband coupling, and $\sqrt{V_{11}V_{22}}/V_{12} \approx 14$ (where $V_{ij}$ – electron-boson interaction matrix elements), whereas the ratio between density of states for the bands with the small and the large gap, $N_2/N_1$, correspondingly, was roughly of an order. We estimated "solo" BCS-ratio values in a hypothetic case of zero interband coupling ($V_{i \neq j} = 0$) for each condensate as $2\Delta_{L,S}/k_B T_C^{L,S} \leq 4.5$. The values are constant within the range of critical temperatures studied, and correspond to a case of strong intraband electron-phonon coupling.

## 1. Introduction

The family of oxypnictide compounds with LnOFePn-type (Ln – lanthanide, Pn - pnictide) structure, or "1111" [1], is the most numerous and covers all the range of critical temperatures, up to 56 K [2], among iron-based superconductors.

Oxypnictides have pronounced layered structure: superconducting Fe-As blocks separating by Ln-O spacers form a "sandwich" along the *c*-direction. Superconductivity emerges by spin density wave state suppression under doping or ambient pressure [3], which is similar to that in cuprates. The most interesting feature of oxypnictides is their multiband nature. Band structure calculations showed hole and electron sheets coexisting at the Fermi surface, where at least two quasi-two-dimensional superconducting condensates form at low temperatures ($T \leq T_C$) [4-6]. To date, the unsolved questions are those concerning mechanism of superconductivity and symmetry type of superconducting gaps [7]. On the one hand, one cannot neglect the coinciding between vector of Fermi surface nesting along Γ-M-direction and vector of antiferromagnetic ground state, which causes a peak of dynamic spin susceptibility ("magnetic resonance" with energy less than the large superconducting gap, $E_{res} \leq 2\Delta_L$) [8]. The latter may hint at possible Cooper pairing on spin fluctuations, which suggests sign-changed order parameter (so called $s^{\pm}$- model) [9,10]. However, experimental data showing the susceptibility peak for iron-based superconductors are rather controversial, whereas the observed peak being smeared do not satisfy the resonance condition $E_{res} \leq 2\Delta_L$ [6, 11]. As for 1111 system, the magnetic resonance was traced only in two works [12, 13] with $E_{res} \approx 2\Delta_L$ [14]. On the other hand, it is possible to describe iron-based superconducting system in framework of orbital fluctuation coupling (constant-sign $s^{++}$-model), where the resonance peak is irrelevant to mechanism of superconductivity [15]. The latter was confirmed in recent work on cuprates [16], where the resonance peak position was shown to be $2\Delta_L$. Despite the popularity of $s^{\pm}$-theory, recently a lot of works supporting $s^{++}$-mechanism appeared [17-21]. Both of the competing theories converge on the electron-phonon origin of intraband coupling in each condensate, which is verified by a strong isotope effect on atoms of iron [22].

The difficulty emerged is that the experimental data on determination of superconducting gaps (to be compared with magnetic resonance energy) and their temperature dependences (which could facilitate estimation of electron-boson coupling constants) are inconsistent. Measurements by the most popular tunnel technique, point contact spectroscopy, give tremendous scatter in the BCS-ratio values for both, the large gap ($2\Delta_L/k_BT_C = 3 - 22$), and the small gap ($2\Delta_S/k_BT_C = 1.7 - 6.8$) ([6, 23-30], see also Tables 1 in [31, 32]). For example, the results obtained in [24, 29] on $SmO_{1-x}F_xFeAs$ samples synthesized similarly to those studied here, differ drastically. Authors of [29] observed Andreev peculiarities in NS-spectra from two superconducting gaps with $2\Delta_{L,S}/k_BT_C$ ratios of 7 - 9 and 2.5 - 3, correspondingly. In contrast, scanning tunneling microscopy studies [24] were able to detect the only gap with the BCS-ratio of about 3.6. At the same time, in [26] NS-Andreev spectroscopy of $SmO_{1-x}F_xFeAs$ gave for the large gap the value $2\Delta_L/k_BT_C \approx 4.7$, for the small gap – 1.7. In [27, 28], a BCS-like temperature behavior of the only gap with the BCS-ratio closed to weak coupling limit 3.52 was observed. Optical studies [30], in contrast, yielded higher BCS-ratio value for the large gap, $2\Delta_L/k_BT_C \approx 8$. Comprehensive studies of $SmFe_{1-x}Ru_xAsO_{0.85}F_{0.15}$ samples with different ruthenium concentrations and the range of $T_C = (13 - 52)$ K [25] unambiguously set two gap peculiarities in NS-spectra and BCS-like gap temperature dependences typical for a strong interband interaction. Summarizing different experimental data on iron-based superconductors, authors of [25] also plotted the dependences of the $\Delta_L$ and $\Delta_S$ characteristic BCS-ratios on $T_C$. According to [25], $2\Delta_{L,S}/k_BT_C$ rises drastically with decreasing of critical temperature, which is, in their opinion, a sign of the coupling mediated by spin fluctuations.

Here we present dynamic conductance spectra of Andreev break-junctions formed in $Sm_{1-x}Th_xOFeAs$ polycrystalline samples (with thorium doping) and critical temperatures

$T_C$ = (34 - 45) K. Using Andreev spectroscopy and intrinsic Andreev spectroscopy we determine the large and the small superconducting gaps and their temperature dependences. We estimate BCS-ratios, anisotropy degree for the large gap, relative electron-boson coupling constants, and the ratio between densities of states for the two bands. A strong electron-phonon coupling is shown to be driven in both condensates of 1111 system, while the interband interaction is weak. Any variation of spacer structure was concluded not to change the mechanism underlying superconductivity, but to influence on the Fermi level density of states. The latter, in turn, leads to $T_C$ variation.

## 2. Experimental details

We used $Sm_{1-x}Th_xOFeAs$ polycrystalline samples with various doping level (0.15 ≤ x ≤ 0.3) synthesized under high-pressure; for details see [33, 34]. X-ray diffraction studies showed a presence of the only superconducting phase $Sm_{1-x}Th_xOFeAs$, while the amount of accidental nonsuperconducting phases $ThO_2$ and SmAs did not exceed 10%. Bulk critical temperatures determined from resistive measurements of the samples were $T_C^{bulk}$ = 45 ± 1 K (NZ5 sample), and $T_C^{bulk}$ = 40 ± 1 K (NZ7, NZ9 samples).

SNS-Andreev spectroscopy being the experimental technique we used for studies of superconducting properties is based on the effect of multiple Andreev reflections from NS-interfaces of superconductor – normal metal – superconductor structure, or SNS [35]. A necessary condition to observe the effect is a small contact diameter *a* which should be less than quasiparticle mean free path *l* [36]. Then a set of peculiarities at certain bias voltages $V_n = 2\Delta/en$, so-called subharmonic gap structure (SGS), would appear in I(V)- and dI(V)/dV-characteristics of SNS-contact [37-39]. In a case of high transparency contact, series of dynamic conductance minima would be observed [39, 40]. A presence of two such structures corresponding to each gap is typical for two-gap superconductor. Due to finite value of *l*/*a* ratio, probability of elastic penetration of SNS-interface by a quasiparticle decreases at low bias voltages. This leads to both, fading of relative amplitude of Andreev peculiarities with their number increasing, and the presence of an excess current area, so-called foot, in current-voltage characteristic (CVC) appearing at $V_n \to 0$. Thus, the main signs to multiple Andreev reflections effect are the foot in I(V)-characteristic within the range of low bias voltages, and the SGS in dynamic conductance spectrum of the contact. The latter manifestation, in accordance with theory [39], facilitates determination of superconducting order parameter using Andreev reflex positions as $2\Delta = \langle eV_n \cdot n \rangle$ at any temperature within the range of $0 < T \leq T_C$. Therefore, Andreev spectroscopy of symmetrical SNS-contacts provides a unique opportunity to direct measurements of superconducting gaps and their temperature dependences without any dynamic conductance spectrum fitting; the latter is required, for example, for unsymmetrical NS-contacts.

To implement multiple Andreev reflections spectroscopy experimentally, we exploited the "break-junction" technique [41]. Corners of superconductor sample, which was prepared as thin rectangular plate of about 3×1.5×0.2 mm$^3$, were covered by indium-gallium solder spots; in such the way the sample was fixed on a spring sample holder to provide 4-contact connection. Then the sample holder was cooled down to T = 4.2 K and was bended mechanically. This led to the sample natural cleavage with a formation of contact of two superconducting banks through a weak link (ScS-contact, where $c$ is constriction). Obviously, the microcrack was generated remotely from current and potential leads, which excluded an overheating of the contact area during the current applying, and guaranteed a good heat sink. Precise curving of the sample holder implies both, providing of a controlled sliding of two superconducting banks apart, and preventing a cryogenic cleft degradation.

CVC and dynamic conductance spectra presented here are typical for classical SNS-contact of high transparency [39]. The weak link thus formally acts as normal metal. In terms of average values of mean free path $l \approx 13$ nm [42, 43], bulk resistivity of the samples in the normal state $\rho_n \approx 8 \times 10^{-5}$ Ω·cm [34], and resistance of the contacts studied $\langle R \rangle \approx 50$ Ω, one can estimate the contact diameter [36] as $a = \sqrt{4\rho_n l/(3\pi R)} \approx 10$ nm. This value is of the same order as the quasiparticle mean free path. Therefore, it is possible to obtain ballistic break-junctions and to observe multiple Andreev reflections effect using these samples. Number of Andreev peculiarities is expected to be 1–2.

It is widely known that steps-and-terraces are formed on clefts of any layered compounds (including polycrystalline ones). In superconducting oxypnictides they can be realized as S-N-S-N-…-S-contacts, where Fe-As layers play a role of superconductors, spacers act as weak links. Intrinsic multiple Andreev reflections effect (IMARE, firstly reported in [44]) being similar to intrinsic Josephson effect [45, 46] was observed on such contacts in cuprates and, later, on other layered superconductors. As a height of each step is divisible by $c$ lattice parameter, such junctions being identical are electrically equivalent to the single ScS-junctions connected in series. Therefore, bias voltage of any CVC peculiarities depicting *bulk* properties of the material would be $N$ times greater than that in single ScS-junction characteristic. Bias voltage of Andreev reflexes for such array will hence coincide with $V_n$ of minima in single-junction characteristic under normalizing, while the superconducting gap value could be determined as $2\Delta \cdot N = \langle eV_n \cdot n \rangle$. As the stack is in fact a part of natural structure of the material, a defect contribution (caused by the formation of surfaces – cryogenic clefts) to the dynamic conductance decreases with $N$ increasing. The latter becomes clear in the experiment: according to our data [47], the increasing of N makes SGS minima in spectra of Andreev-type stacks sharper, which boosts the accuracy of evaluation of the order parameter essentially, and allows concluding on anisotropy degree. With it, probing of namely intrinsic multiple Andreev reflections effect guaranties determination of superconducting order parameters in a bulk. A mechanical readjustment providing easily by the "break-junction" technique makes it possible to form tens of single contacts and arrays on cryogenic clefts of one and the same sample.

As estimated in [34], transport properties of $Sm_{1-x}Th_xOFeAs$ polycrystalline samples we used are strongly anisotropic (anisotropy degree $\gamma \approx 10$), which means a pronounced layered structure. The high values of critical fields $H_{c1}$ are followed from qualitative intergrain boundaries. As the contact diameter estimated is much less than both, crystallite dimensions (about 70 mcm [34]), and typical terrace width (50-500 nm [47]), this opens a unique possibility to use the "break-junction" technique for symmetrical contact forming not only in single crystals, but in polycrystalline samples of layered compounds too [47], as well as to local probing of the superconducting properties.

Thus, the superconducting order parameters were probed by the two methods: Andreev spectroscopy of SNS-contacts (MARE), and intrinsic Andreev spectroscopy of S-N-S-N-…-S-type stacks (IMARE). Both methods provide direct and local (i.e. for point contact with diameter of about tens nanometers) determination of gap values at any temperatures till $T_C$. At the same time, IMARE spectroscopy is more preferable because guaranties measuring of namely bulk properties and excluding of low-quality contacts, such as grain-grain ones.

### 3. Experimental results and discussion

Spectra of SNS-Andreev contacts obtained on cryogenic clefts of nearly optimally doped $Sm_{1-x}Th_xOFeAs$ polycrystalline sample (labeled as NZ5) with bulk critical temperature $T_C^{bulk} \approx 45$ K are presented in Fig. 1. Upper characteristic depicts dynamic conductance of single SNS-contact #d1 obtained in NZ5 sample at T = 4.2 K (further, we will refer to the contact as NZ5_d1 for brevity). Dynamic conductance rises dramatically at low bias voltages being the so called foot area, which is typical for classical SNS-Andreev contact. Two series of Andreev peculiarities are clear: for the large gap, at $V_{L1} \approx \pm 22$ mV and $V_{L2} \approx \pm 11.9$ mV, as well as for the small gap, at $V_{S1} \approx \pm 6$ mV and $V_{S2} \approx \pm 3.2$ mV. According to SGS formula, the positions of these dips define the value of two superconducting gaps: $\Delta_L \approx 11.5$ meV (that gives BCS-ratio $2\Delta_L/k_BT_C^{bulk} \approx 5.9$), and $\Delta_S \approx 3.1$ meV. When assuming this spectrum to correspond to $N$-junction stack (where $N \geq 2$), the BCS-ratio thus would be $2\Delta_L/k_BT_C^{bulk} \approx 5.9/N \leq 3$, which is obviously impossible for driving band because the value must exceed the weak-coupling BCS-limit 3.52. Therefore, to determine gap values for spectra of the arrays obtained consequently on one and the same sample, one should normalize the spectra by $N$ times in accordance with the single-junction spectrum (NZ5_d1). Fig. 1 also shows normalized dynamic conductance for stack contacts NZ5_d2 (number of junctions in the array N = 2), NZ5_d3 (N = 4), NZ5_d4 (N = 6), NZ5_d6 (N = 5), and NZ5_d8 (N = 7). The dI(V)/dV-curves were shifted vertically for clarity. When scaling the bias voltage axis for these spectra by 2, 4, 6, 5, and 7, correspondingly, we achieved the coincidence between positions of Andreev peculiarities for both, the large gap (the positions are marked by gray areas covering 10% uncertainty, and by $n_L$ labels), and the small gap (the range of values is marked by dashed areas, the corresponding peculiarities are signed by arrows and $n_S$ labels). Note that bias voltages of the gap peculiarities observed at SNS-Andreev spectra are integer multiple by $2\Delta_i$, which would be impossible in case of grain-grain contacts.

Therefore, we reproducibly observe multiple Andreev reflections effect and intrinsic multiple Andreev reflections effect, those manifested as two independent SGSs in SNS-contact spectra. In addition, it is worth noting the gap values to be self-consistent when determined by both the methods. This means a good quality of cryogenic clefts obtained. Moreover, taking into account the position of peculiarities caused by adverse influence of surface to be independent on $N$ (see above), one can conclude on bulk nature of both, the large and the small gaps.

To determine the large and the small gap values, we plot the dependence of Andreev dips position $V_n$ on their reversed number $1/n$ (Fig. 2). Solid data points in the figure correspond to the large gap peculiarities, open data points match to the small gap ones. As expected in accordance with formula $V_n = 2\Delta/en$, the experimental data are well-approximated by two straight lines of different slope, both crossing (0,0) point. Hence, the data do form two independent SGSs. The average value of the large gap ($T_C^{local} \approx 45$ K) is thus $\Delta_L = 10.8 \pm 1.1$ meV, of the small gap – $\Delta_S = 2.9 \pm 0.4$ meV. Relatively large dispersion of the small gap values for these spectra may be caused by anisotropy as well as the location of the $\Delta_S$-minima peculiarities on the high-conductance area (foot) corresponding to the band with the large gap, which obstructs the determination of the peculiarity positions. The BCS-ratios can be estimated as $2\Delta_L/k_BT_C^{local} \approx 5.6 >> 3.52$, and $2\Delta_S/k_BT_C^{local} \approx 1.5 << 3.52$.

The fine structure of Andreev minima for $\Delta_L$ also is worth-noting: the doublet-shape of the first peculiarity is clear in spectra for #d1, #d3, #d4, #d6, and #d8 contacts. This may point to the $\Delta_L$ gap anisotropy within the range from 10% to 30% (the positions of the minima in the doublet determine the minimal and the maximal value of the order parameter). The deviation from "pure" s-wave symmetry is also concluded from relatively wide Andreev minima.

Fig. 3 shows normalized spectra for contacts with low critical temperatures ($T_C^{local} \approx 37$ K) obtained on $Sm_{0.85}Th_{0.15}OFeAs$ samples at T = 4.2 K: NZ9_d8 (N = 6), NZ7_d17 (N = 6), NZ7_c (N = 6), NZ7_d (N = 6), and NZ9_d14 (N = 8). For comparison, we show normalized CVC with excess current typical for SNS-Andreev mode, for NZ7_c contact. As in Fig. 1, the positions of SGS minima for both gaps coincide well after the normalization of these spectra by corresponding integer $N$. The most intensive minima located at $V_{L1} \approx \pm 16.6$ mV and $V_{S1} \approx \pm 3.4$ mV, in accordance with the aforementioned formula determine $2\Delta_L$ and $2\Delta_S$, correspondingly. Andreev peculiarities of higher order ($n_{L,S} = 2$) located at $V_{L2} \approx \pm 8.3$ mV and $V_{S2} \approx \pm 1.7$ mV are clear only in spectra of the most qualitative contacts which diameter is seemed to be minimal. A fine structure of the main minima for the large gap is well-reproducible. Dashed line in Fig. 3 shows the position of additional peculiarity $V^* \approx \pm 12.5$ mV which is caused by bulk properties of the material and can be interpreted as exhibiting the k-space anisotropy of the large gap. This means the maximal value of the driving order parameter to be about $\Delta_L^{max} = eV_{L1}/2 \approx 8.3$ meV, and the minimal value to be about $\Delta_L^{min} = eV^*/2 \approx 6.3$ meV, whereas the anisotropy degree is of the order of 25% (extended s-wave symmetry). The $V_n(1/n)$-dependence for spectra from Fig. 3 is shown in Fig. 4. Similarly to that for high-$T_C$ contacts, the positions of the Andreev minima observed groups into two straight lines. Their slopes give the average gap values: the large gap $\Delta_L = 8.3 \pm 0.7$ meV, and the small gap $\Delta_S = 1.7 \pm 0.2$ meV ($T_C^{local} \approx 37$ K). Interestingly, the BCS-ratios for each gap are still constant herewith: $2\Delta_L/k_BT_C^{local} \approx 5.2 \gg 3.52$, and $2\Delta_S/k_BT_C^{local} \approx 1.1 \ll 3.52$.

To obtain the temperature dependence for the large and the small gap in nearly optimal doped NZ5 sample, we measured the dynamic conductance spectrum of NZ5_c contact (N = 4) within the temperature range 4.2 K ≤ T ≤ 45 K, till the $T_C$ (Fig. 5). The characteristics were shifted vertically along with temperature raise for clarity. In the spectrum measured at T = 4.2 K, the positions of Andreev dips for the large gap $\Delta_L \approx 10.5$ meV are marked by $n_L$ labels, for the small gap $\Delta_S \approx 2.8$ meV – by arrows and $n_S$ labels. According to the theory [39], temperature smearing affects only the relative amplitude rather than the position of Andreev minima in spectrum of symmetrical SNS-contact. Explicitly, the minima corresponding to both gaps move toward the low biases and become less intensive with the temperature increasing; finally, the spectrum appears to be linear. The latter means the transition of the contact area (of about 10 nm in diameter) to the normal state, which occurs at the local critical temperature of the contact $T_C^{local} \approx 45$ K.

Using the data presented in Fig. 5, we plot the temperature dependence of the large gap (solid circles), and the small gap (open circles) shown in Fig. 6. Firstly, it is worth comparing their temperature behavior. Crossed circles depict the small gap dependence which was normalized to the large gap value at T = 4.2 K, while dash-dot lines show ordinary single-gap BCS-like function. Evidently, while the large gap dependence $\Delta_L(T)$ hardly deviates from the BCS-type, the $\Delta_S(T)$ dependence differs drastically from $\Delta_L(T)$, thus supporting that peculiarities observed are concluded to describe the properties of different superconducting condensates in the bulk. Moreover, the behavior of the small gap is obviously non-described by single-gap BCS-model. To interpret the phenomenon, we fitted the temperature dependences $\Delta_{L,S}(T)$ by two-gap model by Moskalenko and Suhl [48,49] with renormalized BCS-integral (the fitting is presented in Fig. 6 by gray solid lines). According to this model, the shape of temperature dependences depends on values of four electron-boson coupling constants $\lambda_{ij} = V_{ij} \cdot N_j$ ($V_{ij}$ – matrix interaction element, $N_j$ – density of quasiparticle states at Fermi level in the normal metallic state for $j^{th}$ band), where two of them are intraband ($i = j$), and other two are interband ($i \neq j$). The temperature dependences obtained are typical for weak interband coupling: at low temperatures the small gap closes drastically, while the large gap "sags" relatively to the BCS-like function. Further, while $\Delta_S(T)$ fades to zero, $\Delta_L(T)$ returns to the standard single-gap dependence, then tends to $T_C$ abruptly.

Similar dependences were obtained and plotted for NZ7_c contact with low critical temperature ($T_C^{local} \approx 37$ K). The spectrum measured within 4.2 K $\leq$ T $\leq$ $T_C^{local}$ is shown in Fig. 7. Here the dynamic conductance curves are also shifted along the vertical scale in order of temperature increasing. The dashed line depicts the spectrum of this contact measured at T = 4.2 K after thermocycling (lower curve). As one can see, the shape and the position of SGS peculiarities in this dI/dV-characteristic are reproducible, which demonstrates a mechanical stability of our break-junction. Andreev dips for the large gap $\Delta_L \approx 8.3$ meV are marked by $n_L$ labels in spectra measured at T = 4.2 K, for the small gap $\Delta_S \approx 1.8$ meV – by $n_S$ labels. Temperature dependences $\Delta_{L,S}(T)$ for this contact are presented in Fig. 8 (solid circles and open point-centered circles, correspondingly). Similarly to the contact with high $T_C^{local}$ (see Fig. 6), temperature increasing affects the positions of Andreev minima caused by $\Delta_L$ and $\Delta_S$ differently. Normalized dependence of the small gap, $\Delta_S(T)\Delta_L(4.2K)/\Delta_S(4.2K)$ marked in Fig. 8 by crossed circles deviates from the large gap dependence. The behavior of both gaps is analogous to that presented in Fig. 6. It deviates typically from single-band BCS-like dependences (shown by dash-dot lines) and fully agrees with Moskalenko and Suhl system of equations [48,49] (the corresponding fitting is presented by gray solid lines). Similar $\Delta_{L,S}(T)$ behavior was observed by us earlier in $Mg_{1-x}Al_xB_2$ [50], LiFeAs [51], and $GdFeAsO_{0.88}$ [52].

The fundamental parameters of superconducting state of $Sm_{1-x}Th_xOFeAs$ determined with a help of fitting by two-gap equation system (see Figs. 6, 8) are summarized in the table. In order to compare, we also present the data on $GdFeAsO_{0.88}$ being another Fe-based oxyarsenide with critical temperature $T_C^{local} \approx 49$ K; the corresponding temperature dependences and their fitting by Moskalenko and Suhl model were published earlier in [52]. Let us compare the values of relative $\lambda_{ij}$ constants (normalized by $\lambda_{11}$). We judge the main role in superconductivity in both, $Sm_{1-x}Th_xOFeAs$ and $GdFeAsO_{0.88}$, to be played by intraband Cooper pairing: along the $T_C^{local}$ variation, the $\lambda_{LL}$ and $\lambda_{SS}$ constants have maximal values (see the table). The ratio $\beta = \sqrt{\lambda_{LL}\lambda_{SS}/\lambda_{LS}\lambda_{SL}} = \sqrt{V_{LL}V_{SS}}/V_{LS}$ demonstrates that the intraband coupling is more than order of magnitude stronger than the interband coupling.

Now we compare the "solo" parameters $T_C^i$ and $2\Delta_i/k_B T_C^i$ (where $i = L,S$) for the two condensates, describing the properties of each of them in a hypothetical case of zero interband interaction. "Solo" BCS-ratios remain unchanged within the range of critical temperatures 37-49 K: both gaps scale with $T_C$. Herewith, for the large gap, the characteristic ratio is about $2\Delta_L/k_B T_C^L = 4.5$-$4.8$. For the small gap, the ratio is averagely lower, $2\Delta_S/k_B T_C^S = 3.8$-$4.5$, albeit is close to that for the $\Delta_L$-condensate. Basing on some experimental data [19,22], one can confidently enough conclude on a strong electron-phonon nature of *intraband* coupling being in the framework of the theory by Eliashberg. The latter also agrees with some recent theoretical calculations [5,7,9,10,15] for both, $s^{++}$ and $s^{\pm}$-model.

So, the values of $\lambda_{SS}/\lambda_{LL}$ as well as "solo" ratios $2\Delta_i/k_B T_C^i$ estimated remain nearly constant in spite of both, the variation of doping concentration in Sm-based samples and underlying lanthanide Sm/Gd, i.e. in the samples studied only chemical composition of spacers rather than Fe-As blocks is changed. The Fe-As blocks have nearly constant degree of structural ordering, whereas only doping level, and, hence, $N_{L,S}$ values are varied. Therefore, spacers in 1111 structure act as charge reservoirs not taking any part in superconductivity directly. Specified variations of composition are seemed to be unaffecting the pairing mechanism and the strength of electron-boson interaction. Basing on it and taking into account both, quasi-two-dimensionality being the same for the $\Delta_L$ and $\Delta_S$ condensates and similar structure of the Fermi surfaces, one can easily explain the observed scaling between each gap and critical temperature. We hence can compare superconducting properties of oxypnictides of different composition using two relative parameters: the ratio of nondiagonal coupling constants (or, which is the same, the ratio between densities of states in the bands $\alpha = \lambda_{LS}/\lambda_{SL} = N_S/N_L$), and the ratio between effective intraband coupling and effective interband one $\beta = \sqrt{V_{LL}V_{SS}}/V_{LS}$. According to our data, these parameters are also still unchanged within the range of $T_C = (37$-$49)$ K: $\alpha \approx 10$ with

10% uncertainty, and $\beta$ = 10-14. High $\beta$ value shows that the condensates interact weakly with each other, whereas the interband interaction is non-negligible when describing such the two-gap system, due to the difference between the densities of states in the two bands.

Turning to the temperature dependences of the gaps (see Figs. 6, 8, and Fig. 2 in [52]), one can conclude that the "tail" observed experimentally in $\Delta_S(T)$ originates from an induced superconductivity in the bands with the small gap by dominating $\Delta_L$-condensate at the temperatures higher than "solo" $T_C^S \sim T_C^{local}/3$ (see the table) through a k-space proximity effect [53]. With it, namely sizable density of states in $\Delta_S$-band (high $\alpha$ value) distorts the temperature dependence of the large gap. As a result, we observe the powerful curvature of $\Delta_L(T)$-dependence in relation to single-gap BCS-like function. Our data also point to the local critical temperature to be up to 20% lower than "solo" $T_C^L$ for condensate owing to the large gap.

In conclusion, "solo" BCS-ratios $2\Delta_i/k_B T_C^i$ for the two bands are close and lie in the range of 4.3 ± 0.5 within $T_C$ = (37-49) K. Such high values of critical temperatures result from a strong *intraband* electron-phonon coupling in the condensate with the large gap. The nature of the bands with the small gap is similar, but the corresponding coupling is weaker, whereas the density of states is of the order of magnitude higher. Variation in chemical composition of spacers in oxypnictides does not affect the mechanism of superconductivity within this $T_C$ range and behaves similar to variation of oxygen atom amount in its effect on the electron subsystem. This probably leads to changing of density of states at the Fermi level in each band, which causes in turn the scaling between both gaps and $T_C$. The conclusion agrees well with that of theoretical work [54].

We thank Ya.G. Ponomarev and V.M. Pudalov for valuable discussions and equipment provided. The study was supported by RFBR Grants #12-02-31269-mol_a, and #13-02-01451-a.

Parameters of two-gap superconducting state of $Sm_{1-x}Th_xOFeAs$ determined directly from SNS-Andreev spectra (Figs. 6, 8)[*)] and determined from the gap temperature dependence fitting using Moskalenko and Suhl model [48,49][**)]

| contact | NZ7_c | NZ5_c | KHL8_f (Gd-1111) [52] |
|---|---|---|---|
| $T_C^{local}$; $T_C^L$; $T_C^S$, K | 37; 43.2; 9.7 | 45; 53.7; 14.4 | 49; 60.3; 18.3 |
| $\Delta_L$, $\Delta_S$, meV | 8.3; 1.8 | 10.5; 2.8 | 12.5; 3.0 |
| $\frac{2\Delta_L}{k_B T_C^{local}}$; $\frac{2\Delta_S}{k_B T_C^{local}}$ | 5.2; 1.1 | 5.4; 1.4 | 6.0; 1.4 |
| $\frac{2\Delta_L}{k_B T_C^L}$; $\frac{2\Delta_S}{k_B T_C^S}$ | 4.5; 4.3 | 4.5; 4.5 | 4.8; 3.8 |
| $T_C^{local}/T_C^L$ | 0.86 | 0.84 | 0.81 |
| $\frac{\lambda_{SS}}{\lambda_{LL}}$; $\frac{\lambda_{LS}}{\lambda_{LL}}$; $\frac{\lambda_{SL}}{\lambda_{LL}}$ | 0.643; 0.178; 0.018 | 0.67; 0.183; 0.018 | 0.628; 0.254; 0.023 |
| $\alpha = \frac{N_S}{N_L}$; $\beta = \frac{\sqrt{V_{LL}V_{SS}}}{V_{LS}}$ | 9.7; 14.0 | 10.3; 14.4 | 11.2; 10.4 |

[*)] Local critical temperature of the contact $T_C^{local}$, value of the order parameters $\Delta_L$, $\Delta_S$ at T = 4.2 K, and corresponding BCS-ratios for the local critical temperature $T_C^{local}$.
[**)] "Solo" BCS-ratios for each condensate $2\Delta_i/k_B T_C^i$ (excluding interband interaction) determined using "solo" critical temperature $T_C^{L,S}$ for the condensates, the ratio $T_C^{local}/T_C^L$, relative coupling constants $\lambda_{ij}/\lambda_{LL}$, the densities of states in the bands $\alpha = \lambda_{LS}/\lambda_{SL} = N_S/N_L$ and the ratio between intraband and interband coupling $\beta = (\lambda_{LL}\lambda_{SS}/\lambda_{LS}\lambda_{LL})^{1/2} = (V_{LL}V_{SS})^{1/2}/V_{LS}$.
[***)] Data for the contact obtained in Gd-1111 sample with $T_C^{local} \approx 49$ K [52] are presented for comparison.

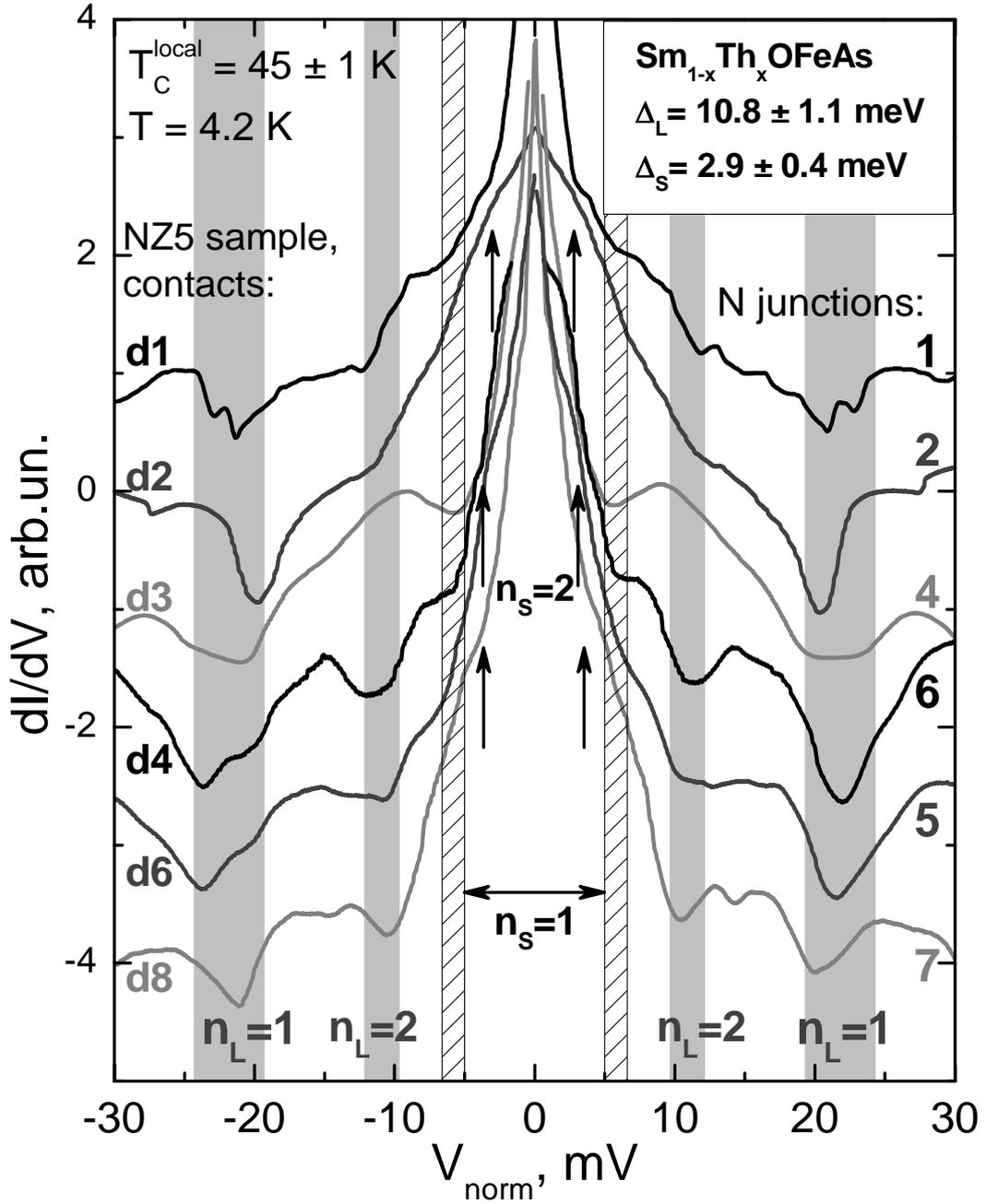

**Fig. 1.** Dynamic conductance of array contacts #d2 (the number of junctions in the stack is $N = 2$), #d3 ($N = 4$), #d4 ($N = 6$), #d6 ($N = 5$) and #d8 ($N = 7$) normalized to the spectrum of single contact #d1. $T_C^{local} \approx 45$ K. All the contacts were obtained on NZ5 sample by sequent mechanical readjustment at $T = 4.2$ K. The position of Andreev dips for the large gap $\Delta_L = (10.8 \pm 1.1)$ meV is marked to by gray areas (covering 10% uncertainty) and $n_L$ labels; for the small gap $\Delta_S = (2.9 \pm 0.4)$ meV – by dashed areas (covering ~15% range of values), arrows and $n_S$ labels.

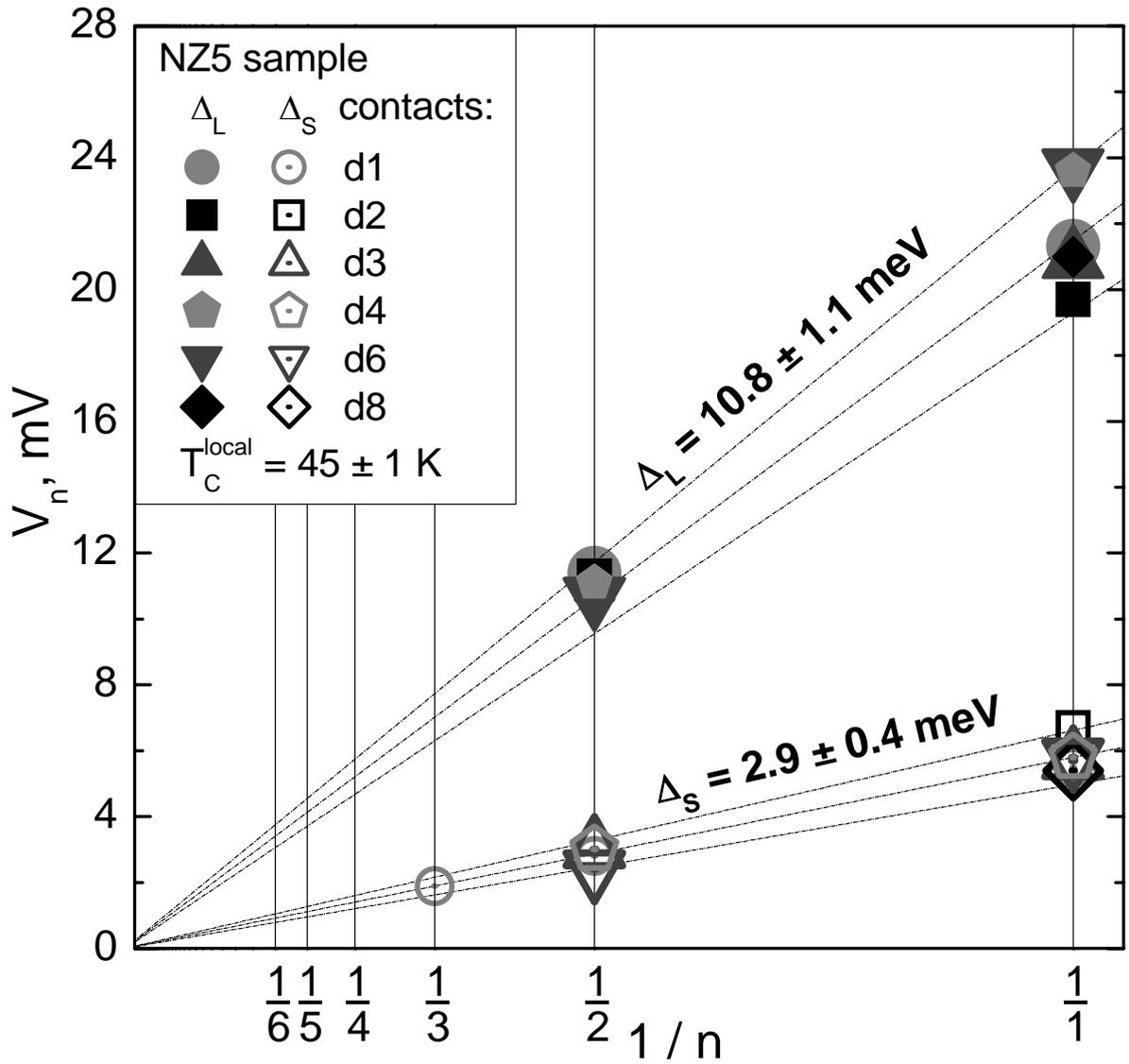

**Fig. 2.** The dependence of Andreev peculiarity positions $V_n$ for the large and the small gap on reversed number $1/n$ for the spectra of contacts with $T_C^{local} \approx 45$ K presented in Fig. 1. Solid data points are related to the large gap, open ones – to the small gap.

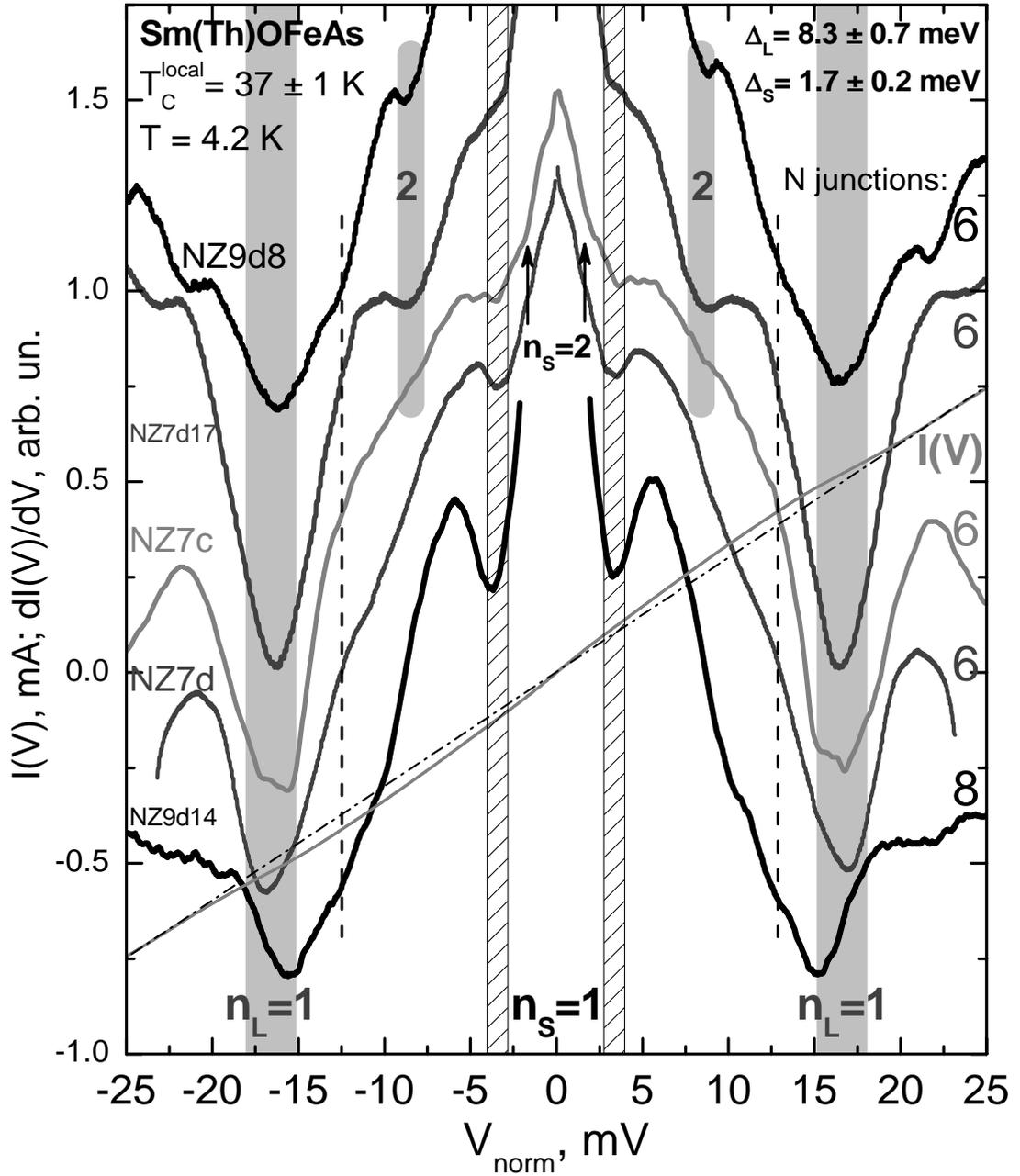

**Fig. 3.** Normalized dynamic conductance of array contacts NZ9_d8 (the number of junctions in the stack is N = 6), NZ7_d17 (N = 6), NZ7_c (N = 6; I(V)-characteristic for this contact is presented by thin gray curve; dash-dot line depicts ohmic dependence), NZ7_d (N = 6) and NZ9_d14 (N = 8). $T_C^{local} \approx 37$ K, T = 4.2 K. The position of Andreev dips for the large gap $\Delta_L = (8.3 \pm 0.7)$ meV is marked to by gray areas (covering 10% uncertainty) and $n_L$ labels; for the small gap $\Delta_L = (1.7 \pm 0.2)$ meV – by dashed areas (covering 10% range of values), arrows and $n_S$ labels. Dashed lines mark the position of the peculiarity relating to anisotropy degree of $\Delta_L$ order parameter.

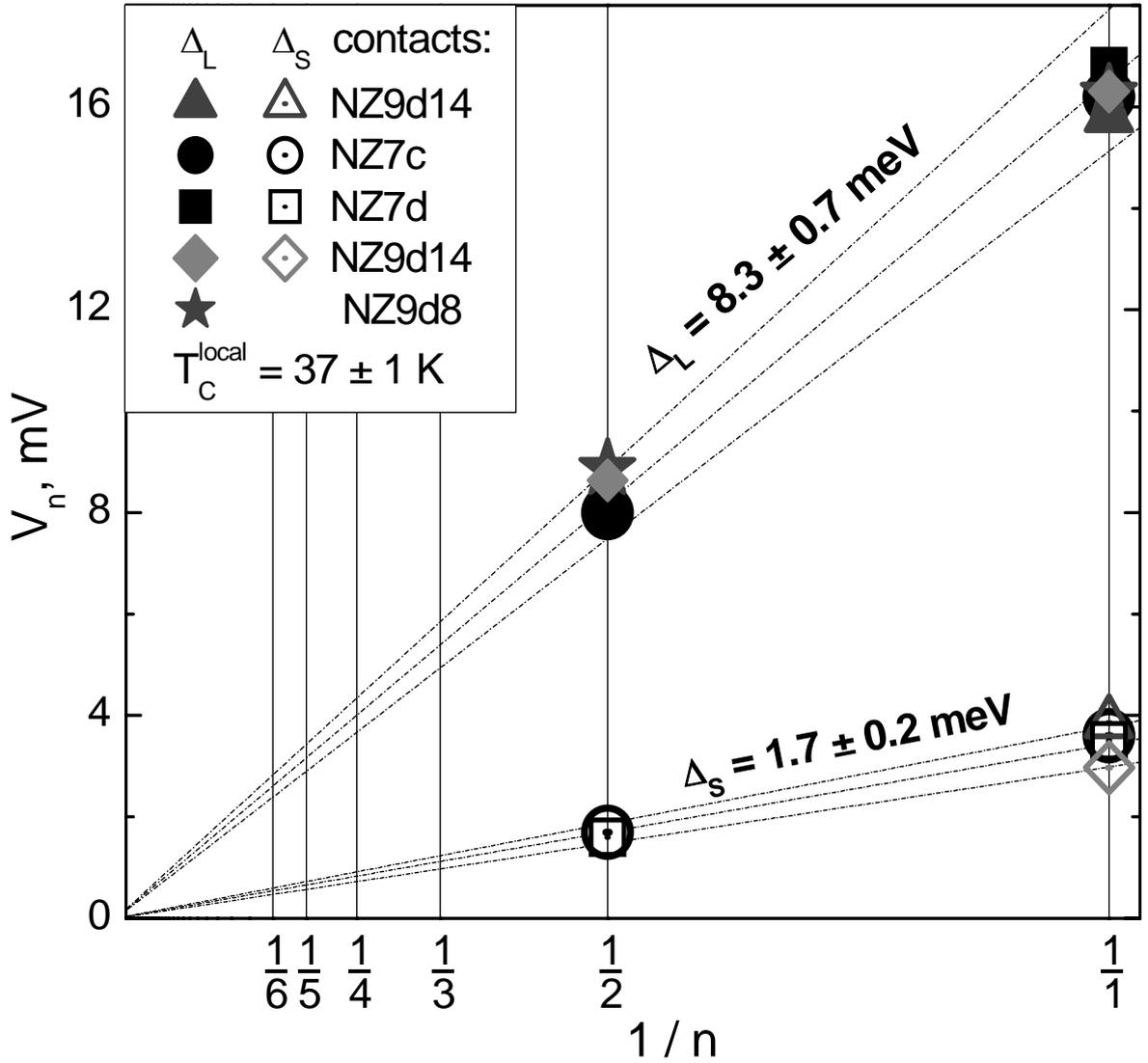

**Fig. 4.** The dependence of Andreev peculiarity positions $V_n$ for the large and the small gap on reversed number $1/n$ for the spectra of contacts with $T_C^{local} \approx 37$ K presented in Fig. 3. Solid data points are related to the large gap, open ones – to the small gap.

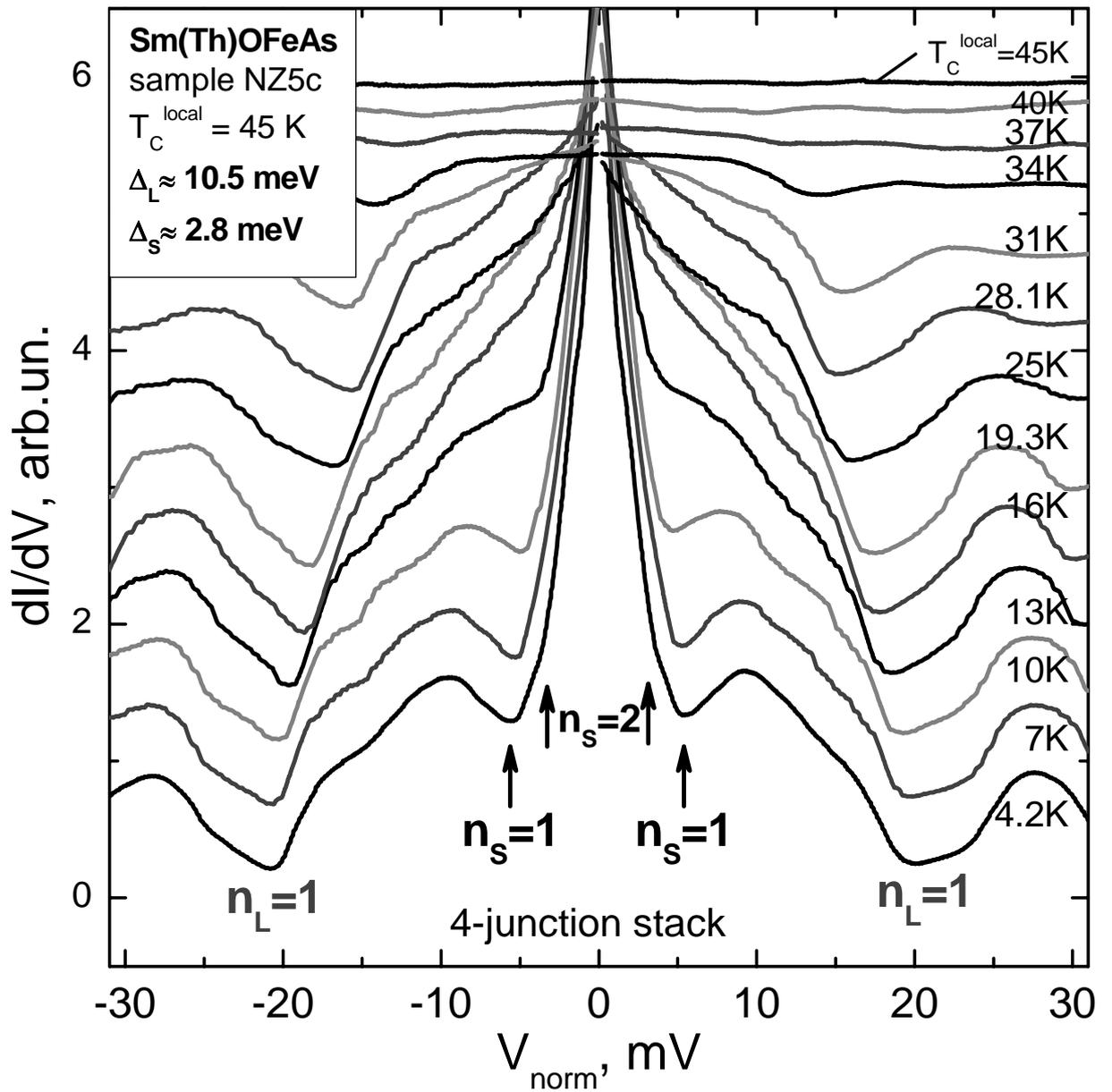

**Fig. 5.** Dynamic conductance spectra of NZ5_c contact measured within the temperature range $4.2\ \text{K} \leq T \leq T_C^{local} \approx 45\ \text{K}$. The position of Andreev dips for the large gap is marked by $n_L$ labels, for the small gap – by arrows and $n_S$ labels.

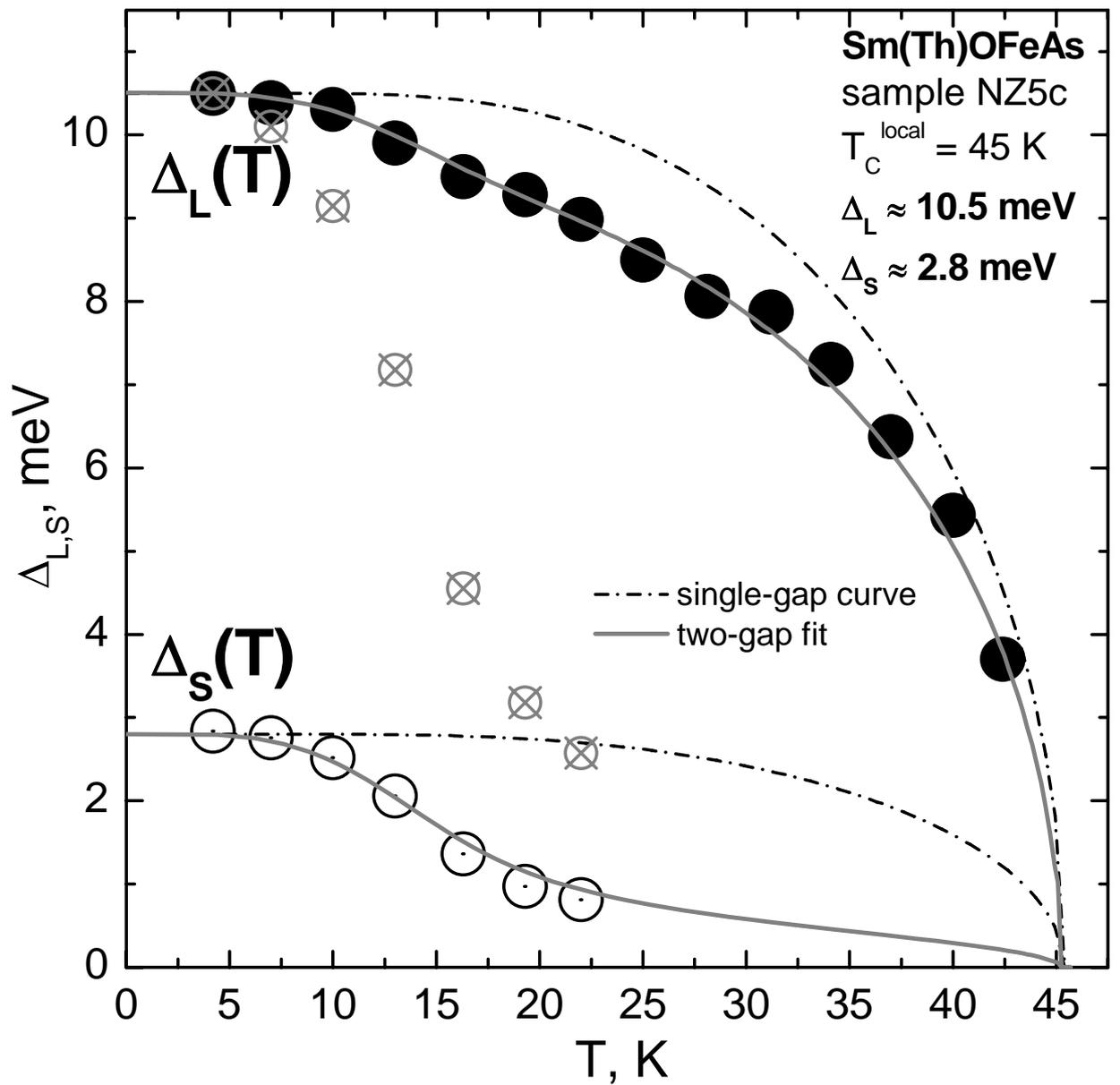

**Fig. 6.** Temperature dependences of the large gap (solid circles) and for the small gap (black open circles) plotted using spectra of the contact with $T_C^{local} \approx 45$ K presented in Fig. 5. Normalized dependence $\Delta_S(T)\cdot\Delta_L(0)/\Delta_S(0)$ is shown by crossed circles for comparison. $\Delta_{L,S}(T)$ fitted by single-gap and two-gap BCS-like models are shown by dash-dot and solid lines, correspondingly.

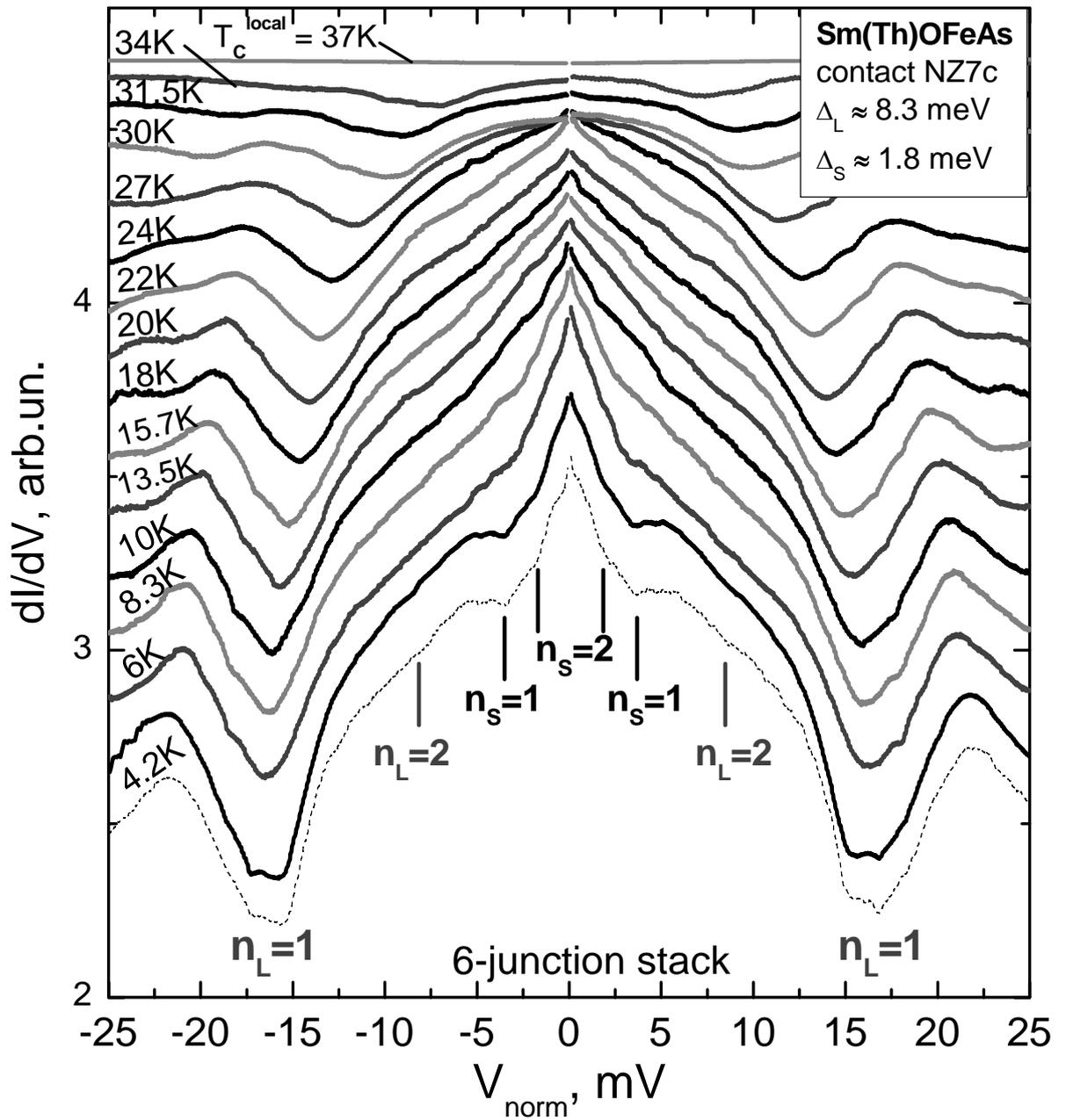

**Fig. 7.** Dynamic conductance spectra of NZ7_c contact measured within the temperature range $4.2\ \text{K} \leq T \leq T_C^{local} \approx 37\ \text{K}$. The position of Andreev dips for the large gap is marked by $n_L$ labels, for the small gap – by $n_S$ labels. Dashed line depicts the dI(V)/dV-characteristic of the contact measured at 4.2 K after thermocycling.

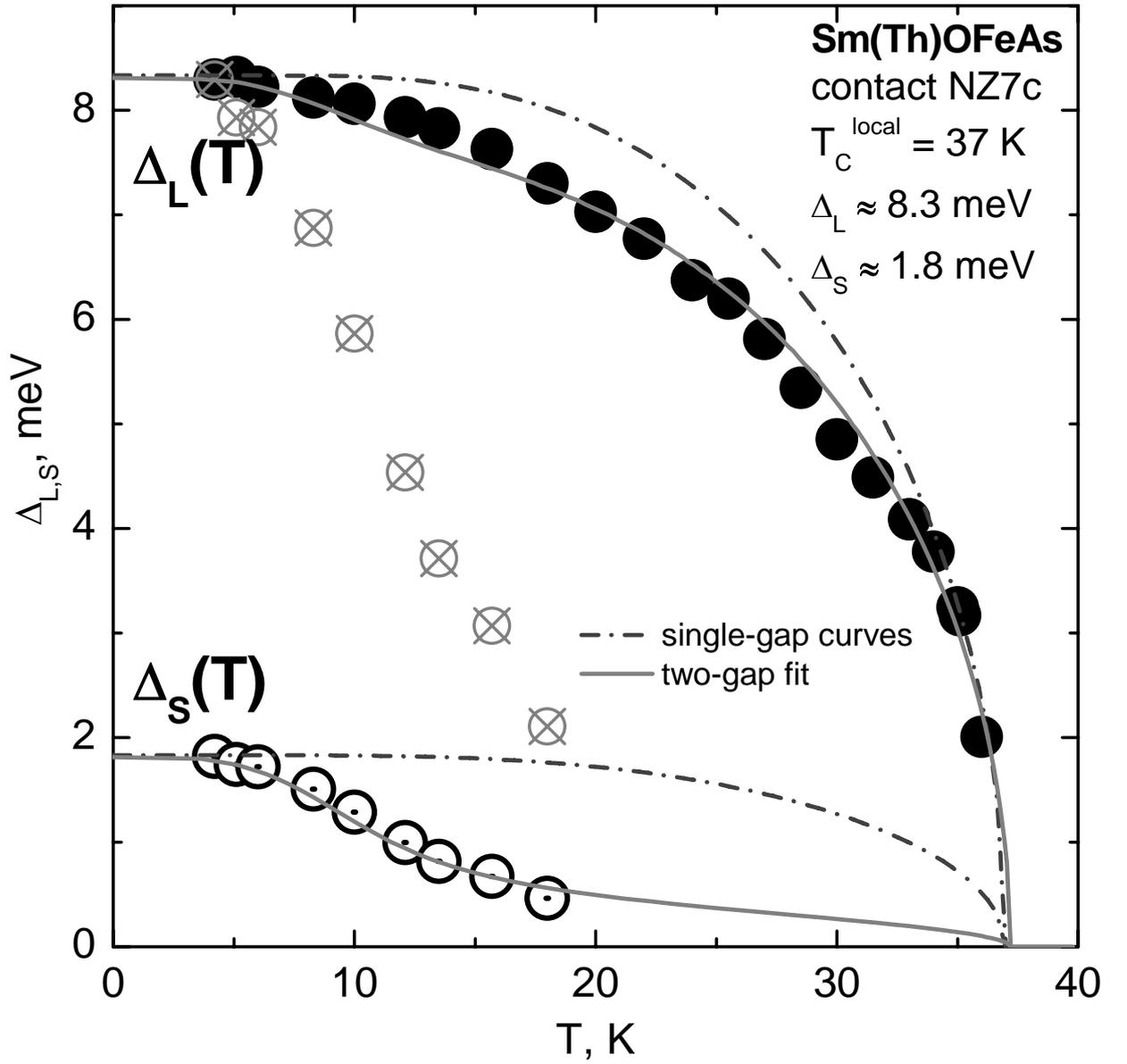

**Fig. 8.** Temperature dependences of the large gap (solid circles) and for the small gap (open circles) plotted using spectra of the contact with $T_C^{local} \approx 37$ K presented in Fig. 7. Normalized dependence $\Delta_S(T) \cdot \Delta_L(0)/\Delta_S(0)$ is shown by crossed circles for comparison. $\Delta_{L,S}(T)$ fitted by single-gap and two-gap BCS-like models are shown by dash-dot and solid lines, correspondingly.

**Literature**
1. Y. Kamihara, T. Watanabe, M. Hirano, H. Hosono, J. Am. Chem. Soc. **130**, 3296, (2008).
2. C. Wang, L. Li, S. Chi, Z. Zhu, Z. Ren, Y. Li, Y. Wang, X. Lin, Y. Luo, S. Jiang, X. Xu, G. Cao, and Z. Xu, Eur. Phys. Lett. **83**, 67006 (2008).
3. H.H. Klauss, H. Luetkens, R. Klingeler, C. Hess, F.J. Litterst, M. Kraken, M.M. Korshunov, I. Eremin, S.L. Drechsler, R. Khasanov, A. Amato, J. Hamann-Borrero, N. Leps, A. Kondrat, G. Behr, J. Werner, and B. Büchner, Phys. Rev. Lett. **101**, 077005 (2008).
4. I.A. Nekrasov, Z.V. Pchelkina, and M.V. Sadovskii, JETP Lett. **87**, 560 (2008).
5. D.J. Singh and M.H. Du, Phys. Rev. Lett. **100**, 237003 (2008).
6. G.R. Stewart, Rev. Mod. Phys. **83**, 1589 (2011).
7. P.J. Hirschfeld, M.M. Korshunov, and I.I. Mazin, Rep. Prog. Phys. **74**, 124508 (2011).
8. C. de la Cruz, Q. Huang, J.W. Lynn, J. Li, W. Ratcliff, J.L. Zarestky, H.A. Mook, G.F. Chen, J.L. Luo, N.L. Wang, and P. Dai, Nature **453**, 899 (2008).
9. I.I. Mazin, D.J. Singh, M.D. Johannes, and M.H. Du, Phys. Rev. Lett. **101**, 057003 (2008).
10. M.M. Korshunov and I. Eremin, Phys. Rev. B **78**, 140509(R) (2008).
11. J. Paglione and R.L. Greene, Nature Phys. **6**, 645 (2010).
12. S. Wakimoto, K. Kodama, M. Ishikado, M. Matsuda, R. Kajimoto, M. Arai, K. Kakurai, F. Esaka, A. Iyo, H. Kito, H. Eisaki, and S. Shamoto, J. Phys. Soc. Jpn. **79**, 074715 (2010).
13. S. Shamoto, M. Ishikado, A.D. Christianson, M.D. Lumsden, S. Wakimoto, K. Kodama, A. Iyo, and M. Arai, Phys. Rev. B **82**, 172508 (2010).
14. Y.G. Ponomarev, S.A. Kuzmichev, T.E. Kuzmicheva, M.G. Mikheev, M.V. Sudakova, S.N. Tchesnokov, O.S. Volkova, A.N. Vasiliev, V.M. Pudalov, A.V. Sadakov, A.S. Usol'tsev, T. Wolf, E.P. Khlybov, and L.F. Kulikova, J. Supercond. Novel Magn. **26**, 2867 (2013).
15. S. Onari and H. Kontani, Phys. Rev. Lett. **104**, 177001 (2009).
16. Y. Zhou, H. Zhang, H. Lin, and C.D. Gong, arXiv:1311.0611 (unpublished).
17. M. Sato, Y. Kobayashi, S.C. Lee, H. Takahashi, E. Satomi, and Y. Miura, J. Phys. Soc. Jpn. **79**, 014710 (2010).
18. S.V. Borisenko, V.B. Zabolotnyy, D.V. Evtushinsky, T.K. Kim, I.V. Morozov, A.N. Yaresko, A.A. Kordyuk, G. Behr, A. Vasiliev, R. Follath, and B. Büchner, Phys. Rev. Lett. **105**, 067002 (2010).
19. K. Ikeuchi, M. Sato, R. Kajimoto, Y. Kobayashi, K. Suzuki, M. Itoh, P. Bourges, A.D. Christianson, H. Nakamura, and M. Machida, arXiv:1310.7424 (unpublished).
20. J. Ishizuka, T. Yamada, Y. Yanagi, and Y. Ono, J. Phys. Soc. Jpn. **82**, 123712 (2013).
21. I.R. Shein and A.L. Ivanovskii, Phys. Lett. A. **375**, 1028 (2011).
22. R.H. Liu, T. Wu, G. Wu, H. Chen, X.F. Wang, Y.L. Xie, J.J. Yin, Y.J. Yan, Q.J. Li, B.C. Shi, W.S. Chu, Z.Y. Wu, and X.H. Chen, Nature **459**, 64 (2009).
23. P. Seidel, Supercond. Sci. Technol. **24**, 043001 (2011).
24. Y. Fasano, I. Maggio-Aprile, N.D. Zhigadlo, S. Katrych, J. Karpinski, and O. Fischer, Phys. Rev. Lett. **105**, 167005 (2010).
25. D. Daghero, M. Tortello, G.A. Ummarino, V.A. Stepanov, F. Bernardini, M. Tropeano, M. Putti, and R.S. Gonnelli, Supercond. Sci. Technol. **25**, 084012 (2012).
26. Y.L. Wang, L. Shan, L. Fang, P. Cheng, C. Ren, and H.H. Wen, Supercond. Sci. Technol. **22**, 015018 (2009).
27. T.Y. Chen, Z. Tesanovic, R.H. Liu, X.H. Chen, and C.L. Chien, Nature **453**, 1224 (2008).
28. Y.G. Naidyuk, O.E. Kvitnitskaya, I.K. Yanson, G. Fuchs, S. Haindl, M. Kidszun, L. Schultz, and B. Holzapfel, Supercond. Sci. Technol. **24**, 065010 (2010).
29. D. Daghero, M. Tortello, G.A. Ummarino, and R.S. Gonnelli, Rep. Prog. Phys. **74**, 124509 (2011).
30. A. Dubroka, K.W. Kim, M. Roessle, V.K. Malik, R.H. Liu, G. Wu, X.H. Chen, and C. Bernhard, Phys. Rev. Lett. **101**, 097011 (2008).

31. T.E. Shanygina, Ya.G. Ponomarev, S.A. Kuzmichev, M.G. Mikheev, S.N. Tchesnokov, O.E. Omel'yanovskii, A.V. Sadakov, Yu.F. Eltsev, A.S. Dormidontov, V.M. Pudalov, A.S. Usol'tsev, and E.P. Khlybov, JETP Lett. **93**, 94 (2011).
32. K.A. Yates, K. Morrison, J.A. Rodgers, G.B.S Penny, J.W.G. Bos, J.P. Attfield, and L.F. Cohen, New J. Phys. **11**, 025015 (2009).
33. N.D. Zhigadlo, S. Katrych, Z. Bukowski, S. Weyeneth, R. Puzniak, and J. Karpinski, J. Phys.: Condens. Matt. **20**, 342202 (2008).
34. N.D. Zhigadlo, S. Katrych, S. Weyeneth, R. Puzniak, P.J.W. Moll, Z. Bukowski, J. Karpinski, H. Keller, and B. Batlogg, Phys. Rev. B **82**, 064517 (2010).
35. A.F. Andreev, Zh. Exp. Teor. Fiz. **46**, 1823 (1964).
36. Yu.V. Sharvin, Zh. Exp. Teor. Fiz. **48**, 984 (1965).
37. M. Octavio, M. Tinkham, G.E. Blonder, and T.M. Klapwijk, Phys. Rev. B **27**, 6739 (1983).
38. G.B. Arnold, J. Low Temp. Phys. **68**, 1 (1987).
39. R. Kümmel, U. Gunsenheimer, and R. Nikolsky, Phys. Rev B **42**, 3992 (1990).
40. J.C. Cuevas, A. Martin-Rodero, and A.L. Yeyati, Phys. Rev. B **54**, 7366 (1996).
41. J. Moreland and J.W. Ekin, J. Appl. Phys. **58**, 3888 (1985).
42. T. Okada, H. Takahashi, Y. Imai, K. Kitagawa, K. Matsubayashi, Y. Uwatoko, and A. Maeda, Phys. Rev. B **86**, 064516 (2012).
43. Ya.G. Ponomarev, S.A. Kuzmichev, M.G. Mikheev, M.V. Sudakova, S.N. Tchesnokov, O.S. Volkova, A.N. Vasiliev, T. Hanke, C. Hess, G. Behr, R. Klingeler, and B. Büchner, Phys. Rev. B **79**, 224517 (2009).
44. Ya.G. Ponomarev, K.K. Uk, M.A. Lorentz, et al., Inst. Phys. Conf. Ser. **167**, 241 (2000).
45. B.A. Aminov, L.I. Leonyuk, T.E. Oskina, H. Piel, Y.G. Ponomarev, H.T. Rachimov, K. Sethupathi, M.V. Sudakova, and D. Wehler, Adv. Supercond. **V**, 1037 (1993).
46. H. Nakamura, M. Machida, T. Koyama, and N. Hamada, J. Phys. Soc. Jpn. **78**, 123712 (2009).
47. T.E. Kuzmicheva, S.A. Kuzmichev, M.G. Mikheev, Ya.G. Ponomarev, S.N. Tchesnokov, Yu.F. Eltsev, V.M. Pudalov, K.S. Pervakov, A.V. Sadakov, A.S. Usoltsev, E.P. Khlybov, and L.F. Kulikova, Eur. Phys. Lett. **102**, 67006 (2013).
48. V.A. Moskalenko, Fiz. Met. Metall. **8**, 503 (1959).
49. H. Suhl, B.T. Matthias, and L.R. Walker, Phys. Rev. Lett. **3**, 552 (1959).
50. Ya.G. Ponomarev, S.A. Kuzmichev, N.M. Kadomtseva, M.G. Mikheev, M.V. Sudakova, S.N. Chesnokov, E.G. Maksimov, S. I. Krasnosvobodtsev, L.G. Sevast'yanova, K.P. Burdina, and B.M. Bulychev, JETP Lett. **79**, 484 (2004); S.A. Kuzmichev, T.E. Shanygina, S.N. Tchesnokov, and S.I. Krasnosvobodtsev, Solid State Comm. **152**, 119 (2012).
51. S.A. Kuzmichev, T.E. Kuzmicheva, A.I. Boltalin, I.V. Morozov, JETP Lett. **98**, 722 (2013).
52. T.E. Shanygina, S.A. Kuzmichev, M.G. Mikheev, Ya.G. Ponomarev, M.V. Sudakova, S.N. Tchesnokov, Yu.F. Eltsev, V.M. Pudalov, A.V. Sadakov, A.S. Usol'tsev, E.P. Khlybov, and L.F. Kulikova, J. Supercond. Novel Magn. **26**, 2661 (2013).
53. I.K. Yanson, V.V. Fisun, N.L. Bobrov, Yu.G. Naidyuk, W.N. Kang, E.M. Choi, H. J. Kim, and S.I. Lee, Phys. Rev. B **67**, 024517 (2003).
54. E.Z. Kuchinskii, I.A. Nekrasov, and M.V. Sadovskii, JETP Lett. **91**, 518 (2010).